# On inertial-range scaling laws

By JOHN C. BOWMAN†

Inertial-range scaling laws for two- and three-dimensional turbulence are re-examined within a unified framework. A new correction to Kolmogorov's $k^{-5/3}$ scaling is derived for the energy inertial range. A related modification is found to Kraichnan's logarithmically corrected two-dimensional enstrophy-range law that removes its unexpected divergence at the injection wavenumber. The significance of these corrections is illustrated with steady-state energy spectra from recent high-resolution closure computations. Implications for conventional numerical simulations are discussed. These results underscore the asymptotic nature of inertial-range scaling laws.

chao-dyn/9507011   2 Aug 1995

## 1. Introduction

The energy spectrum of fully-developed homogeneous turbulence is thought to be composed of three distinct wavenumber regions: a region of energy injection at the largest scales, an intermediate "inertial range" characterized by zero forcing and zero dissipation, and, at the very smallest scales, a region dominated by viscosity. In 1941, Kolmogorov proposed his famous $k^{-5/3}$ scaling law for the inertial-range energy spectrum of homogeneous and isotropic three-dimensional turbulence. Since then, extensive numerical and experimental scrutiny has essentially confirmed this result. Kolmogorov's argument was extended to the two-dimensional enstrophy range by Kraichnan, who suggested the scaling

$$E(k) \sim k^{-3} \left[\ln\left(\frac{k}{k_1}\right)\right]^{-1/3}, \qquad (1.1)$$

where $k_1$ is the lowest wavenumber in the inertial range (Kraichnan 1971a). However, the true inertial-range behaviour of two-dimensional turbulence is still a subject of much controversy.

Until the recent high-resolution work of Borue (1993), virtually all numerical simulations of two-dimensional Navier-Stokes turbulence have suggested an energy spectrum steeper than $k^{-3}$, often more like $k^{-4}$. Those results conflict not only with Kolmogorov's dimensional reasoning but also with atmospheric observations (Boer & Shepherd 1983) and statistical theories of turbulence. Many researchers attribute this steepening to the presence of coherent structures (McWilliams 1984) since these long-lived formations are mistreated by low-order statistical theories. Santangelo *et al.* (1989) and Benzi *et al.* (1990) have argued that the actual spectral behaviour depends strongly on the initial vorticity distribution.

The present work began with the idea that at least some of the observed steepening might actually be due to the logarithmic correction in (1.1), which has often been ignored by previous researchers. In Fig. 1 we compare the functions $k^{-3}$, $k^{-4}$, and equation (1.1). As pointed out by Herring *et al.* (1974), the logarithmically corrected $k^{-3}$ law can easily be mistaken for a $k^{-4}$ law at low wavenumbers. The logarithmic correction is certainly not negligible near the injection wavenumber; in fact, it diverges at $k = k_1$. This is illustrated in the graphs of the logarithmic slope

$$\frac{d \ln E(k)}{d \ln k} = -3 - \frac{1}{3 \ln k} \qquad (1.2)$$

of $E(k)$ in Fig. 2 for several values of the parameter $N$, the number of inertial-range decades. For most conventional simulations, $N$ is no larger than 2. Since the data from direct simulations tends to be noisy, the slope of the energy spectrum is usually determined from the slope of a tangent line fitted to the data at some point in the middle of the inertial range. The vertical line in Fig. 2 is intended to indicate the effective wavenumber at which the slope would be evaluated by this technique. We suggest that the logarithmic correction could be especially significant in older simulations, where the forcing and dissipation scales were not well separated.

† Institute for Fusion Studies, The University of Texas, Austin, TX 78712, USA

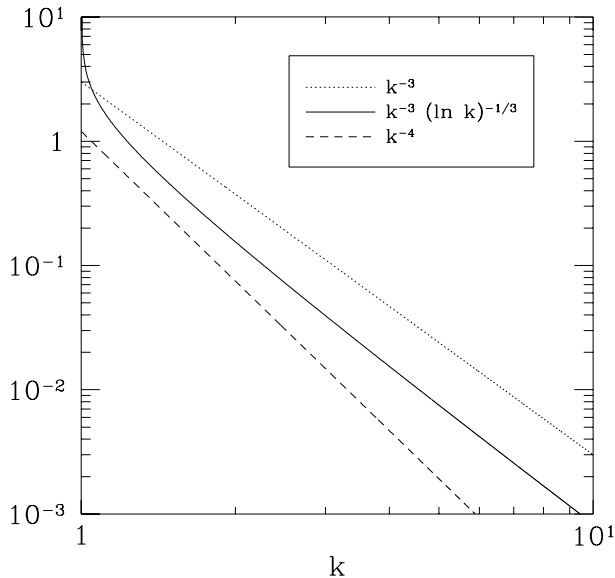

FIGURE 1. Comparison of the scalings $k^{-3}$, $k^{-4}$, and $k^{-3}(\ln k)^{-1/3}$, with arbitrary normalizations.

To add further fuel to this debate, it would be interesting to investigate the predictions of a class of analytical approximations known as statistical closures. These descriptions of turbulence provide approximate evolution equations for the statistical correlation function rather than the velocity field itself. The test-field model [TFM] (Kraichnan 1971b) seems ideally suited for this purpose. Despite the fact that the TFM equations were argued to be dimensionally consistent with (1.1) by Kraichnan (1971a), this has never actually been demonstrated numerically in the literature [*e.g.*, *cf.* Herring (1985)].

## 2. Inertial-range scalings

We begin with a systematic review of the dimensional analysis underlying the Kolmogorov and Kraichnan scalings, focusing on the separate cases of two- and three-dimensional turbulence.

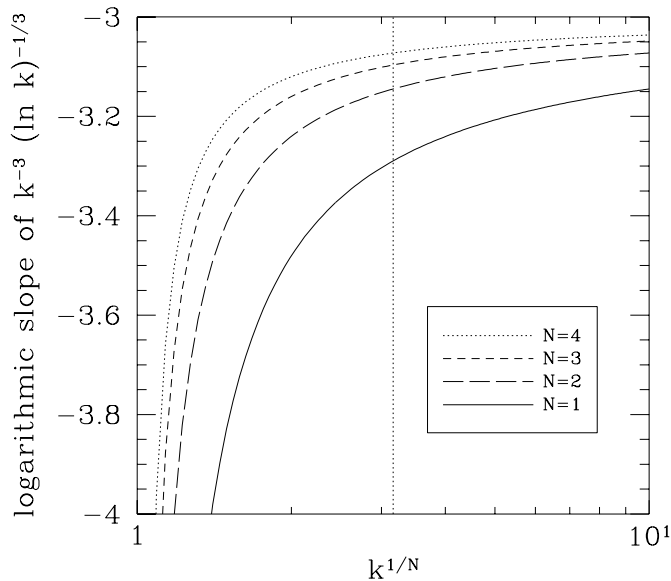

FIGURE 2. Logarithmic slope of the scaling (1.1) for various values of $N$, the number of decades in the inertial range.

### 2.1. *Three-dimensional turbulence*

The Kolmogorov hypothesis relies on the fact that energy is not created or destroyed within the inertial range; it is merely redistributed among the inertial-range wavenumbers. Kolmogorov (1941) suggested that the significant dynamical interactions between the turbulent eddies are *local* in wavenumber space. That is, very large eddies will not interact directly with very small eddies, but only *via* eddies of an intermediate size.

The total energy in all eddies larger than a given scale $k^{-1}$ is $\int_0^k E(\bar{k})\,d\bar{k}$, where $E(k)$ is the energy spectrum. While the shearing effect of the large eddies will significantly distort the small eddies, the random interactions of the many small eddies on the large ones tends to average out their distorting effect. Let us denote the rate of energy transfer to eddies of size $k^{-1}$ and energy $kE(k)$ from larger eddies by $\eta(k)\,kE(k)$, where $\eta(k)$ is the rate at which a unit amount of energy is transferred. Dimensional analysis and the fact that eddies are distorted by the shear in the large-scale flow, rather than by the mean flow itself, leads to the scaling (Kraichnan 1971a)

$$\eta^2(k) \sim \int_0^k \bar{k}^2 E(\bar{k})\,d\bar{k}. \tag{2.1}$$

The rate of energy transfer from eddies larger than $k^{-1}$ to eddies smaller than $k^{-1}$ is hence proportional to the quantity (Ellison 1962)

$$\bar{\Pi}(k) \doteq \left[\int_0^k \bar{k}^2 E(\bar{k})\,d\bar{k}\right]^{1/2} kE(k); \tag{2.2}$$

we will see below that the constant of proportionality is related to the Kolmogorov constant. (We emphasize definitions with the notation '$\doteq$'.)

For statistically stationary turbulence, the amount of energy contained in eddies of a given size is independent of time. Kolmogorov's locality hypothesis would then imply that $\overline{\Pi}$ must be independent of $k$. However, it is well known that real turbulent interactions are not strictly local (particularly in the two-dimensional case discussed below). Indeed, the weighted integral of the energy spectrum appearing in (2.1) actually allows for nonlocal energy transfer. Instead of assuming locality, let us adopt the less restrictive ansatz that, for wavenumbers lying well within the inertial range, the self-similarity of the turbulent interactions makes $\overline{\Pi}(k)$ independent of $k$. We will see in §3 that within the context of statistical closure models, the constancy of $\overline{\Pi}$ is actually a very good approximation, even for the nonlocal two-dimensional enstrophy cascade.

Upon denoting $f(k) = kE(k)$, one may then differentiate the identity

$$\frac{\overline{\Pi}^2}{f^2(k)} = \int_0^k \bar{k} f(\bar{k}) \, d\bar{k} \tag{2.3}$$

with respect to $k$ to determine that $-2\overline{\Pi}^2 f'/f^4 = k$. Integration of this result between some reference wavenumber $k_0$ and $k$ leads to the modified Kolmogorov law

$$E(k) = k^{-1} \left[ \frac{3}{4\overline{\Pi}^2}(k^2 - k_0^2) + k_0^{-3} E^{-3}(k_0) \right]^{-1/3} \qquad (k \geqslant k_0). \tag{2.4}$$

This result may be written more compactly as

$$E(k) = \left(\frac{4}{3}\right)^{1/3} \overline{\Pi}^{2/3} k^{-5/3} \chi^{-1/3}(k) \qquad (k \geqslant k_0), \tag{2.5}$$

in terms of the correction factor

$$\chi(k) \doteq 1 - \frac{k_0^2}{k^2}(1 - \chi_0), \tag{2.6}$$

where $\chi_0 \doteq 4\overline{\Pi}^2 k_0^{-5} E^{-3}(k_0)/3 = \chi(k_0) > 0$. It is often convenient to choose $k_0$ to be the lowest wavenumber in the inertial range.

The correction factor $\chi(k)$ in (2.5) is analogous to the logarithmic correction in Kraichnan's two-dimensional enstrophy cascade law, (1.1). However, (2.5) does not predict a divergence of the energy spectrum at the injection wavenumber since $\chi_0 > 0$. For $k \gg k_0|1 - \chi_0|^{1/2}$, the inertial-range energy spectrum reduces to the usual Kolmogorov law

$$E(k) = \left(\frac{4}{3}\right)^{1/3} \overline{\Pi}^{2/3} k^{-5/3}. \tag{2.7}$$

As was pointed out by Kraichnan (1971a), the dominant contribution to $\eta(k)$ in this limit comes from wavenumbers $\bar{k} \approx k$, as can be seen by substituting (2.7) into (2.1). This is consistent with Kolmogorov's locality hypothesis.

To the author's knowledge the correction factor $\chi(k)$ in (2.5) has not been reported previously. When $\chi_0 \ll 1$, the spectrum will differ from (2.7) only for wavenumbers very close to the injection wavenumber $k_0$, where $\chi(k)$ will lead to a steepening of the energy spectrum, as illustrated in Fig. 3. Notice that the discrepancy is more subtle than in Fig. 2. In the case $\chi_0 \approx 1$, the spectrum will be indistinguishable from (2.7). Finally, in the case $\chi_0 > 1$, there will be a region above $k_0$ over which the spectrum will be less steep than $k^{-5/3}$. We present numerical evidence for this case in §3. In the extreme limit where $k_0 \leqslant k \ll k_0(\chi_0 - 1)^{\frac{1}{2}}$, one expects the energy spectrum to exhibit a $k^{-1}$ behaviour.

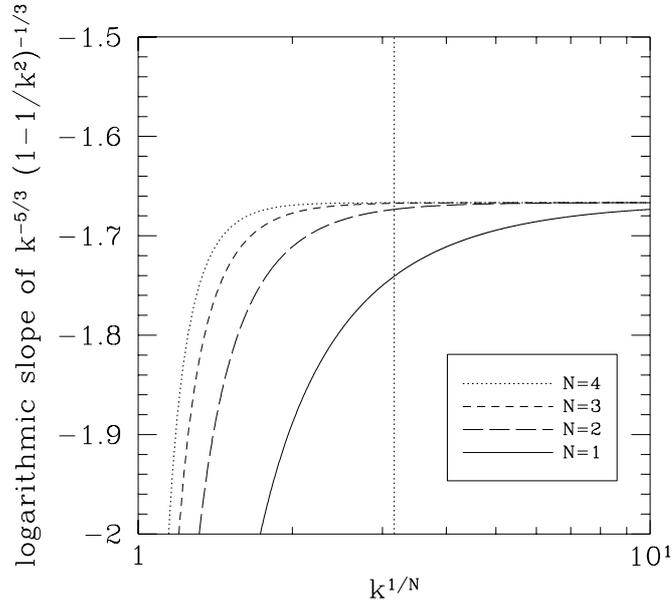

FIGURE 3. Logarithmic slope of the scaling (2.5) for $\chi_0 = 0$ and various values of $N$, the number of decades in the inertial range.

### 2.2. *Two-dimensional turbulence*

Kolmogorov's arguments are based on the conservation of

$$E = \int_0^\infty E(k)\,dk. \tag{2.8}$$

Turbulence in two dimensions is complicated by the presence of an additional *enstrophy* invariant:

$$Z = \int_0^\infty k^2 E(k)\,dk. \tag{2.9}$$

Kolmogorov's picture of energy transfer to the smallest scales cannot be correct in two dimensions since such a redistribution of the energy would imply the creation of new enstrophy (Fjørtoft 1953). Instead, Kraichnan (1967, 1971a) postulated that it is the rate of enstrophy, not energy, transfer that is independent of $k$. The enstrophy transfer rate from eddies larger than $k^{-1}$ to eddies smaller than $k^{-1}$ is proportional to

$$\bar{\bar{\Pi}}_Z(k) \doteq \left[\int_0^k \bar{k}^2 E(\bar{k})\,d\bar{k}\right]^{1/2} k^3 E(k). \tag{2.10}$$

Upon letting $f(k) = k^3 E(k)$ and differentiating as before, we find that $-2\bar{\bar{\Pi}}^2 f'/f^4 = 1/k$. We may integrate this result between some reference wavenumber $k_1$ and $k$ to obtain

$$E(k) = k^{-3}\left[\frac{3}{2\bar{\bar{\Pi}}_Z^2}\ln\left(\frac{k}{k_1}\right) + k_1^{-9} E^{-3}(k_1)\right]^{-1/3} \qquad (k \geqslant k_1). \tag{2.11}$$

It is often convenient to let $k_1$ be the lowest wavenumber in the enstrophy inertial range.

Let us rewrite (2.11) in the form

$$E(k) = \left(\frac{2}{3}\right)^{1/3} \overline{\Pi}_Z^{2/3} k^{-3} \chi^{-1/3}(k) \qquad (k \geqslant k_1), \tag{2.12}$$

where

$$\chi(k) \doteq \ln\left(\frac{k}{k_1}\right) + \chi_1 \tag{2.13}$$

and $\chi_1 \doteq 2\overline{\Pi}_Z^2 k_1^{-9} E^{-3}(k_1)/3 = \chi(k_1)$. Since $\chi_1 > 0$, the divergence exhibited by (1.1) at $k = k_1$ has been removed in (2.12). The logarithmic factor will be significant when $\chi_1 \ll 1$ and for wavenumbers near $k_1$. Upon substitution of (2.12) into (2.1), it is evident that the dominant contribution to $\eta(k)$ is from wavenumbers $\bar{k} \approx k_1$. The enstrophy transfer in two dimensional turbulence is thus seen to be highly nonlocal.

At wavenumbers below $k_1$ an energy inertial range of the form (2.5) will develop, governed by a uniform rate of energy transfer. In this case $k_0$ still represents the lowest wavenumber in the energy inertial range; it is equivalent now not to the highest injection wavenumber but to the highest large-scale dissipation wavenumber. In either two or three dimensions, the eddy distortion (turnover) rate $\eta_k$ for the energy inertial range is given by

$$\eta_k = \left[\int_0^k \bar{k}^2 E(\bar{k}) \, d\bar{k}\right]^{1/2} \sim \frac{1}{kE(k)} \sim \left[k^2 - k_0^2(1 - \chi_0)\right]^{1/3}, \tag{2.14}$$

while for the two-dimensional enstrophy range,

$$\eta_k = \left[\int_0^k \bar{k}^2 E(\bar{k}) \, d\bar{k}\right]^{1/2} \sim \frac{1}{k^3 E(k)} \sim \left[\ln\left(\frac{k}{k_1}\right) + \chi_1\right]^{1/3}. \tag{2.15}$$

### 2.3. *Discussion*

It should be emphasized that the expressions for $\chi(k)$ in (2.6) and (2.13) rely only on the assumption that the quantity $\overline{\Pi}(k)$ defined in (2.2) is independent of $k$. This conjecture is based on the form of $\eta_k$ given in (2.1). One might argue that, while this relation is perhaps valid asymptotically for high $k$, it could miss important large-scale physics and should not be used to determine the form of the energy spectrum near the injection wavenumber. The possibility of new physics entering (2.1) certainly cannot be ruled out; however, the point of the calculation given here is that the self-similarity arguments of Kolmogorov and Kraichnan are actually consistent with scaling relations more general than the classical $k^{-5/3}$ and log-corrected $k^{-3}$ laws.

Even if (2.1) breaks down near the injection wavenumber, (2.5) and (2.12) still contain useful information, provided that (2.1) is approximately valid for wavenumbers $k$ larger than some reference wavenumber $k_0$. In this case, these formulae describe how the asymptotic Kolmogorov–Kraichnan scalings should be matched to the dynamics at scales larger than $k_0^{-1}$.

Eventually, the large-scale corrections proposed in this work should be tested by direct comparison with experiment and numerical simulation. For the time being, the noisiness of experimental and simulation data and the subtlety of the corrections precludes a detailed comparison. However, as a first step towards this goal, we demonstrate in the next section that the constancy of $\overline{\Pi}(k)$ and the resulting corrections to the Kolmogorov theory are at least consistent with the predictions of statistical closure approximations. These pedagogical tools also provide us with a measure of the wavenumber resolution that will be required to verify the proposed modifications directly.

## 3. Closure results

To illustrate the above scalings we will use a recently developed statistical approximation known as the realizable test-field model [RTFM] (Bowman & Krommes 1995). The RTFM is closely related to Kraichnan's TFM but has improved transient behaviour since the random source term in its underlying Langevin representation is not $\delta$-correlated. In the presence of non-Hermitian linear effects (waves) such as those encountered in geophysical and plasma turbulence, the RTFM, unlike the TFM, is guaranteed to predict positive energies (Bowman & Krommes 1995). We will compare the RTFM results to those obtained with the realizable Markovian closure [RMC] (Bowman *et al.* 1993). Like Kraichnan's direct-interaction approximation [DIA] (Kraichnan 1958, 1959, 1961), the RMC is not invariant to random Galilean transformations of the primitive equations (Kraichnan 1964; Leslie 1973); it therefore predicts the incorrect inertial range scaling $k^{-5/2}$ (Herring *et al.* 1974; Bowman 1992). The RMC is closely related to a DIA-based eddy-damped quasinormal Markovian [EDQNM] closure (Orszag 1977; Bowman 1992) but, unlike the EDQNM, it is realizable in the presence of a linear frequency. In a steady state, the RTFM reduces to the TFM equations and the RMC reduces to the EDQNM equations, so that these distinctions need not concern us here.

### 3.1. *Energy spectra*

The closure equations were solved by partitioning the wavenumbers into 64 bins, using the convergent technique of wavenumber partitioning described by Bowman (1994). In Fig. 4 we graph the steady-state energy spectrum for two-dimensional turbulence as predicted by the RTFM closure. (The value 1.0 was chosen for the overall multiplicative factor $g$ entering the expression for the eddy turnover time in the RTFM equations.) To obtain optimal use of the available wavenumber range we replaced the usual Laplacian viscosity $\nu_k = \nu_2 k^2$ with the hyperviscosity $\nu_k = \nu_6 k^6$, where $\nu_6$ was chosen (in terms of $\nu_2$) to keep the enstrophy flux invariant. It was verified that this modification had no effect on the large scale dynamics (Bowman 1994). We estimate the Reynolds number $R = 2\pi (2E)^{1/2}/(k_f \nu_2) \approx 10^{16}$ for this case, with $k_f = 4.25$ and a saturated total energy $E = 5.6$.

The logarithmic slope of the energy spectrum is indicated by the solid line in Fig. 5. We verify in Fig. 6 the linear behaviour of $[k^3 E(k)]^{-3}$ with respect to $\ln(k/k_1)$ as predicted by (2.12), taking $k_1 = 76$. From the slope of the line determined by a least squares fit we calculate $\chi_1 = 3.5$; this value of $\chi_1$ was then used in (2.13) to evaluate the "corrected slope"

$$\frac{d\ln\left[E(k)\,\chi^{1/3}\right]}{d\ln k} \tag{3.1}$$

plotted in Fig. 7. We thus see that an inertial range consistent with (2.12) has developed over about four wavenumber decades. Finally, in Fig. 8 we observe that the corrected eddy distortion rate $\eta_k \chi^{-1/3}$ is nearly constant over the inertial range, in accordance with (2.15).

In contrast, the (DIA-based) RMC closure predicts a slope of $-2.5$, as is illustrated in Fig. 5. As a consequence of its violation of random Galilean invariance, this closure introduces a spurious transfer of enstrophy from large to small scales that leads to an energy spectrum shallower than $k^{-3}$.

By injecting energy at a high wavenumber, $k_f = 1.7 \times 10^7$, and imposing a strict cutoff on the high-wavenumber dissipation, it is possible to focus on the energy inertial range. The steady-state energy spectrum obtained with the RTFM closure is depicted in Fig. 9. In Fig. 12, a region where the logarithmic slope is less than $-5/3$ is apparent near $k = 20$.

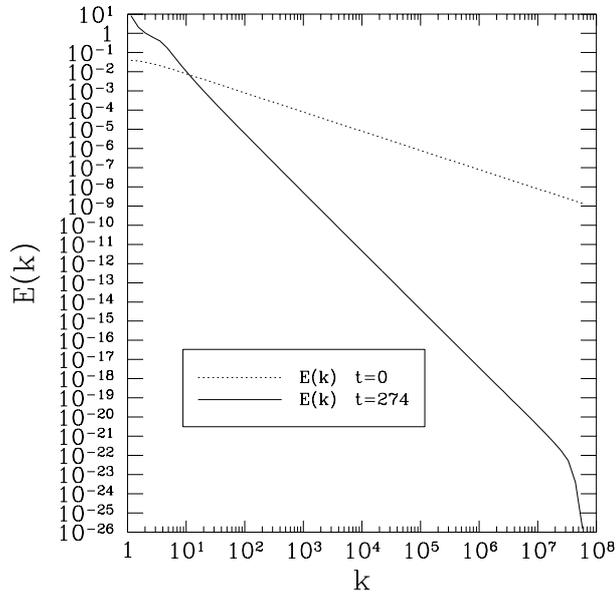

FIGURE 4. Energy spectrum for high Reynolds number two-dimensional fluid turbulence predicted by the RTFM.

This case is an example where $\chi_0 > 1$, as indicated in Fig. 11. A linear least squares fit yields $\chi_0 = 1.13$; this value was used in (2.6) to obtain the corrected logarithmic slope (3.1) shown in Fig. 12. An energy inertial range of the form $k^{-5/3}$ is clearly visible. In Fig. 13 we see that the scaling of the eddy distortion rate is consistent with (2.14).

Finally, if energy is injected at an intermediate wavenumber, $k_f = 3.5 \times 10^4$, both an energy and enstrophy inertial range can develop, as illustrated in Figs. 14 and 15.

### 3.2. Energy and enstrophy transfer

The nonlinear energy transfer function $\Pi_E$ can be defined by

$$\Pi_E(k) \doteq 2 \int_k^\infty d\bar{k}\, T(\bar{k}), \tag{3.2}$$

where $T(k)$ is the triplet correlation function appearing in the energy equation

$$\frac{\partial}{\partial t} E(k) + 2\nu_k E(k) = 2T(k). \tag{3.3}$$

If the nonlinear term is conservative, then

$$\int_0^\infty d\bar{k}\, T(\bar{k}) = 0, \tag{3.4}$$

so that $\Pi_E$ may be equivalently written as

$$\Pi_E(k) = -2 \int_0^k d\bar{k}\, T(\bar{k}). \tag{3.5}$$

Note that (3.4) implies

$$\Pi_E(0) = \Pi_E(\infty) = 0. \tag{3.6}$$

The flow of energy to the high wavenumbers across a surface of constant wavenumber $k$

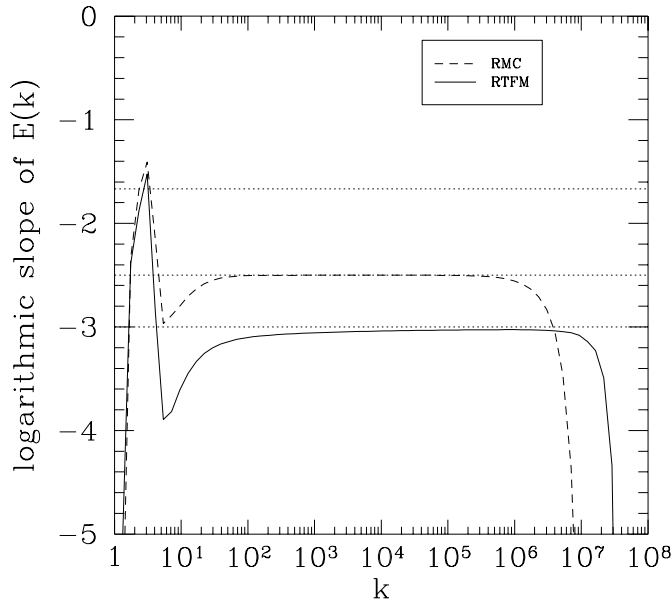

FIGURE 5. Logarithmic slope of the RTFM energy spectrum in Fig. 4 (solid line) and RMC prediction (dashed line).

may then be written in terms of its nonlinear and linear contributions:

$$\frac{\partial}{\partial t}\int_k^\infty d\bar{k}\, E(\bar{k}) = \Pi_E(k) - \epsilon_E(k), \tag{3.7}$$

where $\epsilon_E(k) \doteq 2\int_k^\infty d\bar{k}\, \nu_{\bar{k}} E(\bar{k})$ is the total linear forcing into all wavenumbers higher than $k$. A positive (negative) value for $\Pi_E(k)$ represents a flow of energy to wavenumbers higher (lower) than $k$.

For the two-dimensional inverse energy cascade, one would expect $\Pi_E(k)$ to be negative to the left of the injection range, as is observed in Fig. 16. Since the system is very close to a steady state, the solid and dashed lines, which respectively depict the linear ($\epsilon_E$) and nonlinear ($\Pi_E$) contributions to the energy transfer, coincide. Note that (3.6) is obeyed. At earlier times, one finds that while (3.6) is always satisfied, the linear contribution differs substantially from the nonlinear contribution; this is an indication that the spectrum is still evolving.

In a similar manner, one may define the enstrophy transfer $\Pi_Z$, plotted in Fig. 17. Since it is positive in the enstrophy inertial range, this graph confirms that enstrophy is indeed being transferred to higher wavenumbers.

## 4. Conclusions

This work has highlighted the importance of the logarithmic correction in the enstrophy cascade. Now that the strict divergence in this correction has been removed, the role of this factor should be taken more seriously by the community in comparisons of theoretical scalings with numerical simulation data. The existence of a less significant energy inertial-range correction was also demonstrated in this work. The arguments are applicable even to highly nonlocal turbulence (such as is encountered in two dimensions), provided that there is sufficient self-similarity to make $\bar{\Pi}$ constant within the inertial range.

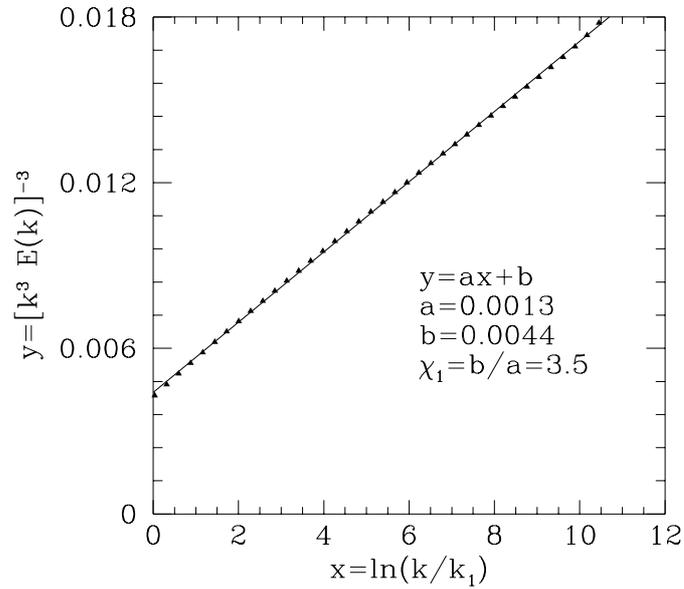

FIGURE 6. Linearity of $[k^3 E(k)]^{-3}$ with respect to $\ln(k/k_1)$ for $k \geqslant k_1 = 76$. The solid triangles are the RTFM predictions.

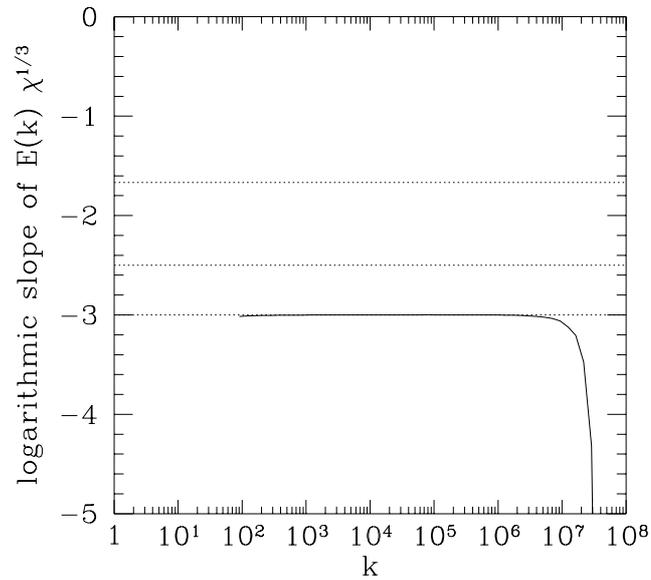

FIGURE 7. Corrected logarithmic slope of the energy spectrum in Fig. 4.

The asymptotic nature of inertial-range scaling laws must be emphasized. The theoretical scalings are expected only in the limit of an infinite inertial range, *i.e.*, where the dissipation and forcing wavenumbers are widely separated. The wavenumbers at the ends of a finite inertial range are influenced by the shape of the energy spectrum outside the inertial range and will not exhibit true inertial-range behaviour. This is especially true for the enstrophy cascade, where the nonlinear transfer is more nonlocal than in the energy range.

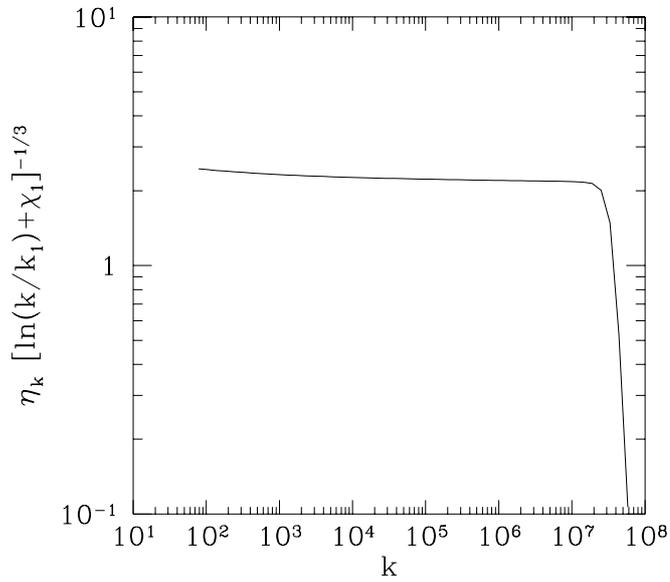

FIGURE 8. Corrected eddy distortion rate $\eta_k \chi^{-1/3}$ for the energy spectrum in Fig. 4.

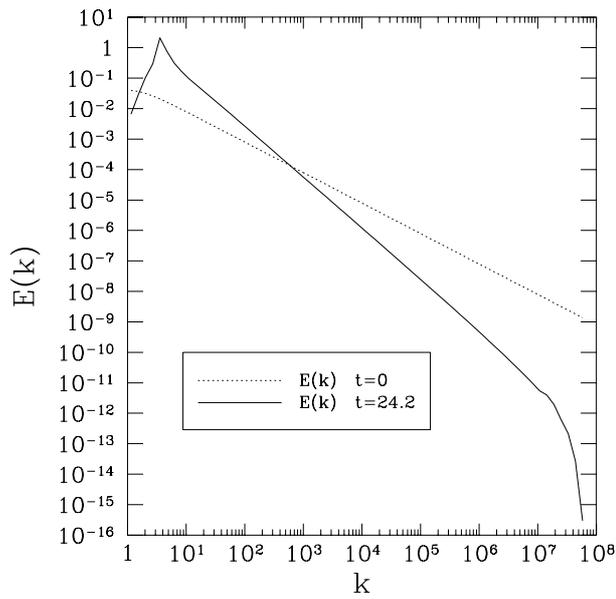

FIGURE 9. Energy inertial range obtained with the RTFM.

In the evaluation of inertial-range exponents, the eye can be easily deceived by the usual guide lines that are drawn tangent to the energy spectrum (*cf.* Fig. 1 and Kraichnan 1991, pp. 76-77). Fortunately, in the case of statistical closure data, it is possible to compute the logarithmic slope of the energy spectrum by exploiting the inherent smoothness of the solutions. One can then gain insight into how widely separated the scales of injection and dissipation must be for a proper inertial range to develop. The numerical results presented in this work suggest that many decades of wavenumber are required. For example, the inertial range that developed in a wavenumber domain of nearly eight decades was only about four decades wide. Even with this much resolution, the theoretical scalings of the eddy turnover times with wavenumber were just barely resolved (*cf.* Figs. 8 and 13).

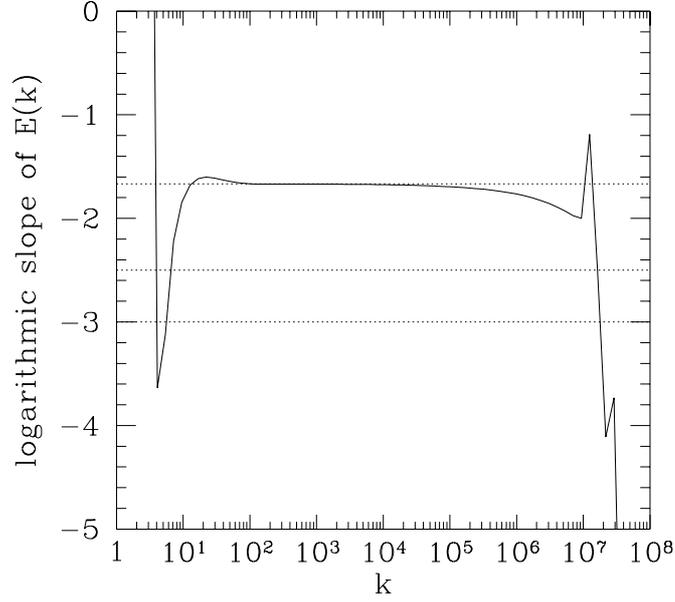

FIGURE 10. Logarithmic slope of the energy spectrum in Fig. 9.

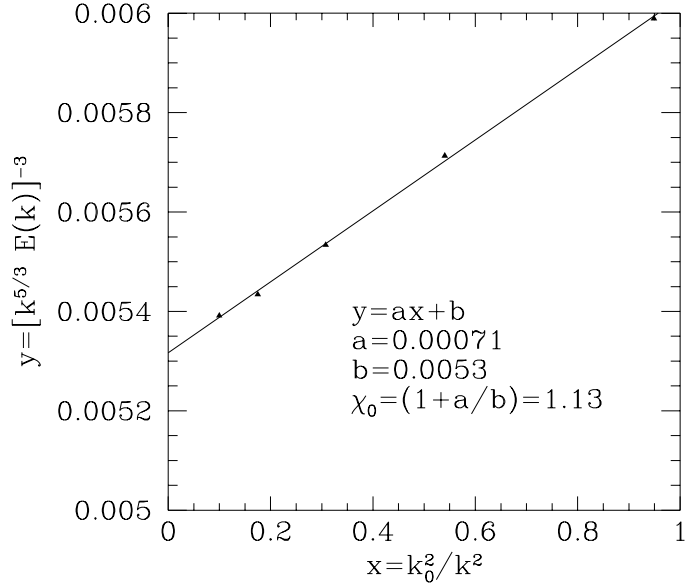

FIGURE 11. Linearity of $[k^{5/3}E(k)]^{-3}$ with respect to $k_0^2/k^2$ for $k \geqslant k_0 = 24.8$. The solid triangles are the RTFM predictions.

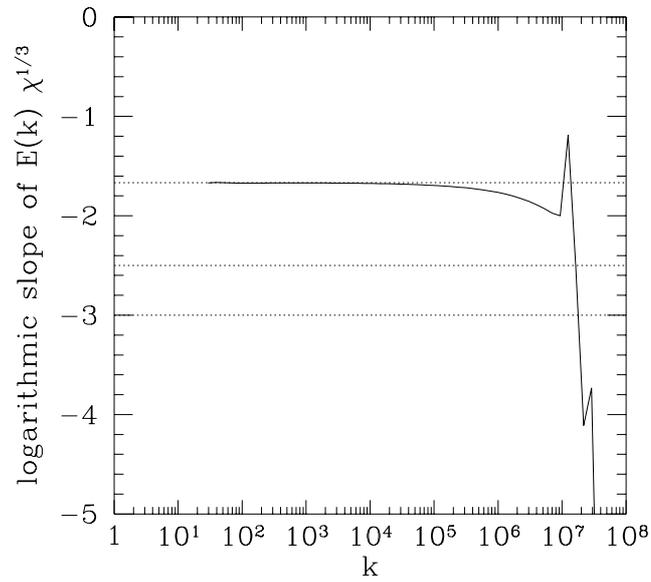

FIGURE 12. Corrected logarithmic slope of the energy spectrum in Fig. 9.

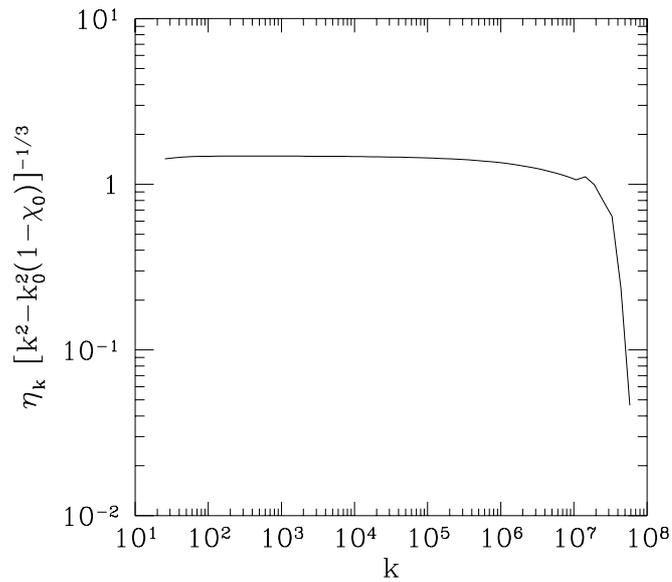

FIGURE 13. Corrected eddy distortion rate $\eta_k \chi^{-1/3}$ for the energy spectrum in Fig. 9.

Given the nonlocality of two-dimensional turbulence, it is not surprising that there has been so much difficulty demonstrating universal behaviour in past conventional simulations of this phenomenon. Perhaps the recent work of Borue (1993) may mark the turning

14    *J. C. Bowman*

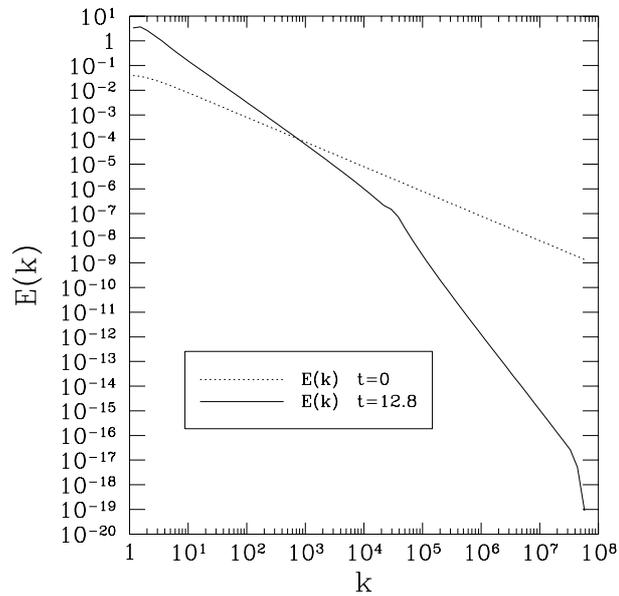

FIGURE 14. Energy and enstrophy inertial ranges obtained with the RTFM.

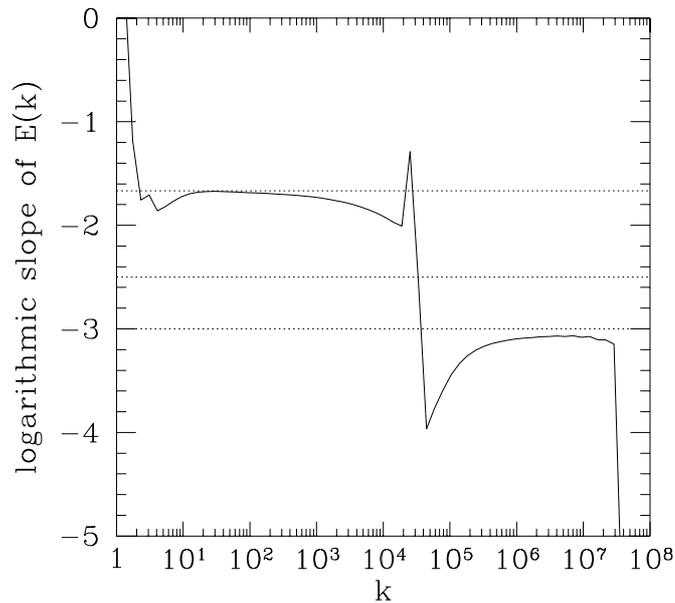

FIGURE 15. Logarithmic slope of the energy spectrum in Fig. 14.

point in this controversy. However, the closure calculations presented here make it clear that very high computer resolution will be required to settle the matter conclusively.

The author is indebted to P. J. Morrison for suggesting the possibility of additional features in the energy inertial range. The author would also like to acknowledge discussions with J. A. Krommes and T. G. Shepherd and financial support from a Natural

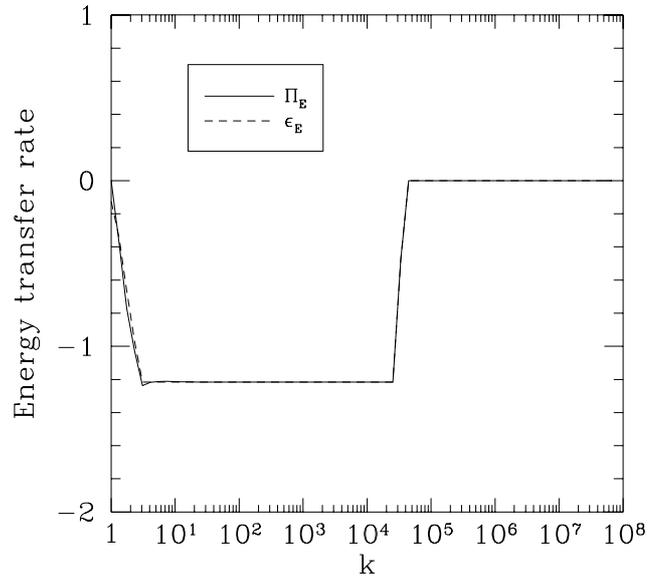

FIGURE 16. Energy transfer function $\Pi_E$ for the energy spectrum in Fig. 14.

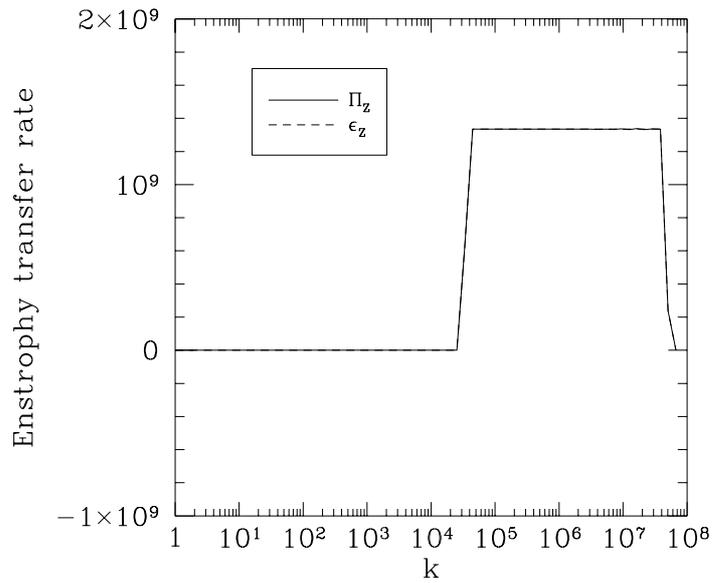

FIGURE 17. Energy transfer function $\Pi_Z$ for the energy spectrum in Fig. 14.

## REFERENCES

BENZI, R., PALADIN, G., & VULPIANI, A. 1990 Power spectra in two-dimensional turbulence.

*Phys. Rev. A* **42**, 3654–3656.

BOER, G. J. & SHEPHERD, T. G. 1983 Large-scale two-dimensional turbulence in the atmosphere. *J. Atmos. Sci.* **40**, 164–184.

BORUE, V. 1993 Spectral exponents of enstrophy cascade in stationary two-dimensional homogeneous turbulence. *Phys. Rev. Lett.* **71**, 3967–3970.

BOWMAN, J. C. & KROMMES, J. A. 1995 The realizable Markovian closure and realizable test-field model. II: Application to anisotropic drift-wave turbulence. To be submitted to *Phys. Fluids B*.

BOWMAN, J. C., KROMMES, J. A., & OTTAVIANI, M. 1993 The realizable Markovian closure. I: General theory, with application to three-wave dynamics. *Phys. Fluids B* **5**, 3558–3589.

BOWMAN, J. C. 1992 *Realizable Markovian Statistical Closures: General Theory and Application to Drift-Wave Turbulence*. Ph.D. thesis, Princeton University Princeton, NJ.

BOWMAN, J. C. 1994 A wavenumber-partitioning scheme for two-dimensional statistical closures. Submitted to *J. Fluid Mech.*

ELLISON, T. H.. "The universal small-scale spectrum of turbulence at high reynolds number" in *Mécanique de la Turbulence* No. 108 pp. 113–121 Paris 1962 C.N.R.S.

FJØRTOFT, R. 1953 On the changes in the spectral distribution of kinetic energy for two dimensional, nondivergent flow. *Tellus* **5**, 225–230.

HERRING, J. R., ORSZAG, S. A., KRAICHNAN, R. H., & FOX, D. G. 1974 Decay of two-dimensional homogenous turbulence. *J. Fluid Mech.* **66**, 417–444.

HERRING, J. R. 1985 Comparison of direct numerical simulation of two-dimensional turbulence with two-point closure: the effects of intermittency. *J. Fluid Mech.* **153**, 229–242.

KOLMOGOROV, A. N. 1941 The local structure of turbulence in incompressible viscous fluid for very large Reynolds numbers. *C. R. Acad. Sci. U.S.S.R.* **30**, 301–306.

KRAICHNAN, R. H. 1958 Irreversible statistical mechanics of incompressible hydromagnetic turbulence. *Phys. Rev.* **109**, 1407–1422.

KRAICHNAN, R. H. 1959 The structure of isotropic turbulence at very high Reynolds numbers. *J. Fluid Mech.* **5**, 497–543.

KRAICHNAN, R. H. 1961 Dynamics of nonlinear stochastic systems. *J. Math. Phys.* **2**, 124–148.

KRAICHNAN, R. H. 1964 Decay of isotropic turbulence in the direct-interaction approximation. *Phys. Fluids* **7**, 1030–1047.

KRAICHNAN, R. H. 1967 Inertial ranges in two-dimensional turbulence. *Phys. Fluids* **10**, 1417–1423.

KRAICHNAN, R. H. 1971 Inertial-range transfer in two- and three-dimensional turbulence. *J. Fluid Mech.* **47**, 525–535.

KRAICHNAN, R. H. 1971 An almost-Markovian Galilean-invariant turbulence model. *J. Fluid Mech.* **47**, 513–524.

KRAICHNAN, R. H. 1991 Turbulent cascade and intermittency growth. *Proc. R. Soc. London, Ser. A* **434**, 65–78.

LESLIE, D. C. 1973 *Developments in the Theory of Turbulence*. Clarendon Press, Oxford.

MCWILLIAMS, J. C. 1984 The emergence of isolated coherent vortices in turbulent flow. *J. Fluid Mech.* **146**, 21–43.

ORSZAG, S. A. 1977 "Lectures on the statistical theory of turbulence" in *Fluid Dynamics*, edited by Balian, R. & Peube, J.-L. pp. 236–373 Gordon and Breach, London (summer school lectures given at grenoble university, 1973).

SANTANGELO, P., BENZI, R., & LEGRAS, B. 1989 The generation of vortices in high-resolution, two-dimensional decaying turbulence and the influence of initial conditions on the breaking of self-similarity. *Phys. Fluids A* **1**, 1027–1034.